\documentclass[a4paper,11pt]{article}
\usepackage{pos}

\title{Theoretical advances in electroweak, Higgs, and top physics at the LHC}
\author*[a]{Mathieu Pellen}

\affiliation[a]{Universit\"at Freiburg, Physikalisches Institut, \\
Hermann-Herder-Str. 3, 79104 Freiburg, Germany}

\emailAdd{mathieu.pellen@physik.uni-freiburg.de}

\abstract{In these proceedings, I review recent precision calculations relevant for the LHC, all related to the electroweak, Higgs or top sector of the Standard Model.
These applications range from triboson production to the production of a top-antitop pair in association with a W boson and to Higgs production in association with another Higgs boson, a top-antitop pair or a W boson.
These proceedings reflect a presentation given at DIS2024 and provide a snapshot of the current frontier for theoretical predictions at the LHC.}

\FullConference{31st International Workshop on Deep Inelastic Scattering (DIS2024)\\
 8–12 April 2024\\
Grenoble, France\\}

\begin{document}
\maketitle

\section{Introduction}

Since its discovery in 2012, the Higgs boson has been one of the main driving forces of the physics programme of the Large Hadron Collider (LHC).
In particular, a lot of emphasis has been put on studying Higgs-boson properties~\cite{Jakobs:2023fxh}.
It is therefore natural that, even after more than 10 years, Higgs physics is still playing an important role at the LHC.

This situation will be reflected in these proceedings by covering several recent state-of-the-art calculations, all related to some extent to Higgs physics at the LHC.
In particular, a combined calculation for triboson and WH production will be highlighted.
Recent progress on electroweak (EW) corrections for double-Higgs production will be shortly reviewed.
Approximate predictions at NNLO QCD for the total cross section of Higgs production in association with a top-antitop pair will be discussed along with progresses regarding the computation of the corresponding two-loop contribution.
Finally, new results for ttW production, which is an important background for ttH measurements, will be shown.

I would like to emphasis that this selection constitutes only a personal and therefore biased view on recent progresses in theoretical calculations for Standard Model predictions at the LHC.
The main motivation for this selection is that it provides a representative overview of the current theory frontier at the LHC.
For more references, the interested reader can look into the Les Houches wishlist of 2021~\cite{Huss:2022ful} which offers an extensive overview of recent theoretical calculations.
This review will be updated in the next few months and a preview can be found in Ref.~\cite{Andersen:2024czj}.
Also, more details on recent calculations for Higgs-related processes at the LHC can be found in Ref.~\cite{Jones:2023uzh}.

\section{Triboson and WH production}

\begin{figure}[h]
\includegraphics[page=1,width=0.32\linewidth]{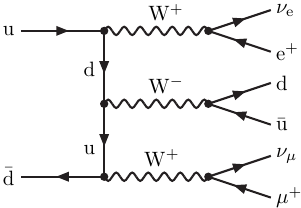}
\includegraphics[page=3,width=0.32\linewidth]{figures/fdiagrams.pdf}
\includegraphics[page=2,width=0.32\linewidth]{figures/fdiagrams.pdf}
\caption{Sample Feynman diagrams for triboson (left and middle) and WH production (right) taken from Ref.~\cite{Denner:2024ufg}.}
\label{fig:diagTriWH}
\end{figure}

In Ref.~\cite{Biedermann:2017bss}, full NLO QCD and EW corrections were computed for both the EW and QCD production of two same-sign W boson in association with two jets which is the golden channel for vector-boson scattering (VBS) at the LHC.
This process being a particularly important probe of the EW sector.
This very final state can actually also be obtained from triboson- and WH-production mechanisms where one W boson is decaying hadronically, as illustrated in Fig.~\ref{fig:diagTriWH}.

\begin{figure}
\includegraphics[width=0.49\linewidth]{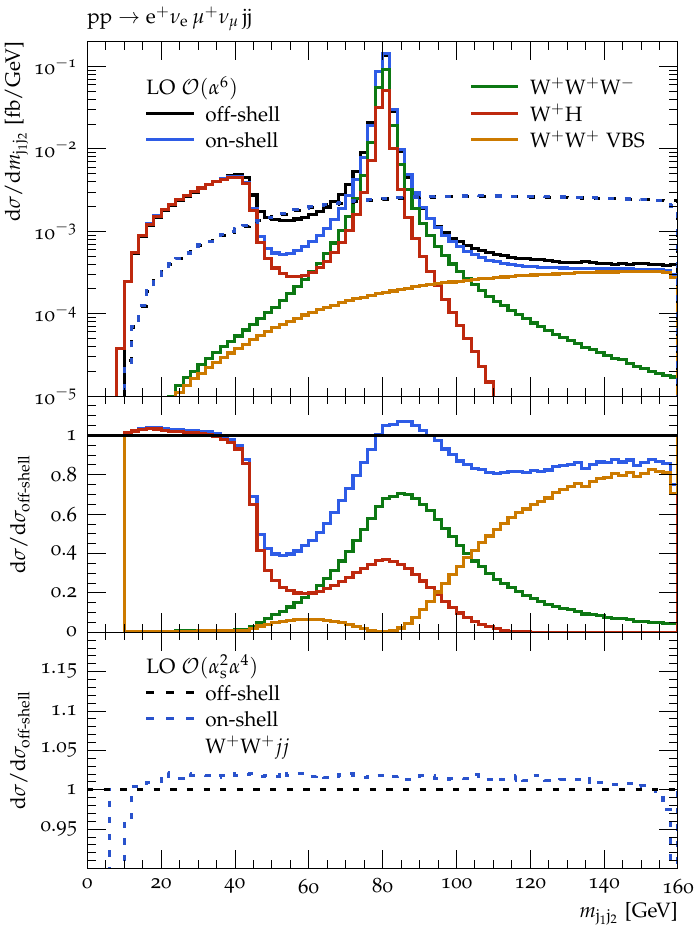}
\includegraphics[width=0.49\linewidth]{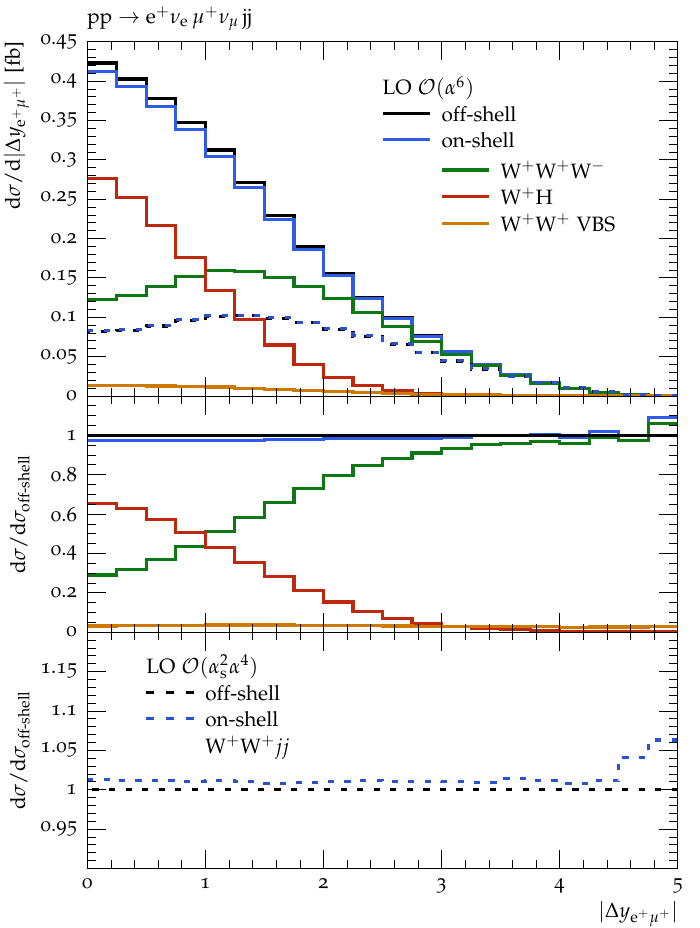}
\caption{Differential predictions at leading-order (LO) for the various production mechanisms compared to the full off-shell calculation for the ${\rm W}^+{\rm W}^+{\rm j}{\rm j}$ signature. The distributions are taken from Ref.~\cite{Denner:2024ufg}.}
\label{fig:LO}
\end{figure}

In Ref.~\cite{Denner:2024ufg}, a new full calculation has been performed for a triboson phase space inspired by a recent ATLAS analysis~\cite{ATLAS:2022xnu}.
The first interesting finding of this work is that the triboson phase space used by the ATLAS collaboration does not only contain triboson production but also WH production at the level of $40\%$ of the fiducial cross section.
This can also be seen in the differential distributions of Fig.~\ref{fig:LO} where the various production mechanisms display a non-trivial interplay across the whole phase space.
Making a distinction between these different mechanisms is therefore somehow artificial.

Beyond the purely phenomenological interest of the calculation, there are also more theoretical questions that motivated this study.
For example, in Ref.~\cite{Biedermann:2016yds}, it has been shown that large EW corrections are an intrinsic feature of VBS at the LHC.
On the other hand, triboson production has shown rather moderate EW corrections.
Given that these two processes share the same final state, these two statements seem contradictory.
Reference~\cite{Denner:2024ufg} has shown that while EW corrections are still large for $t$-channel contributions (like for VBS), they are suppressed with respect to the $s$-channel ones that are enhanced but display small EW corrections.

\begin{figure}[h]
\includegraphics[width=0.49\linewidth]{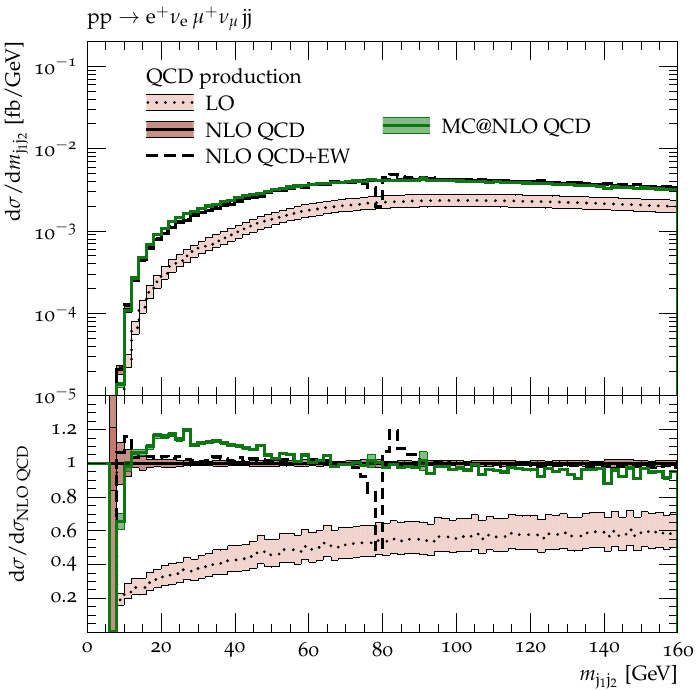}
\includegraphics[width=0.49\linewidth]{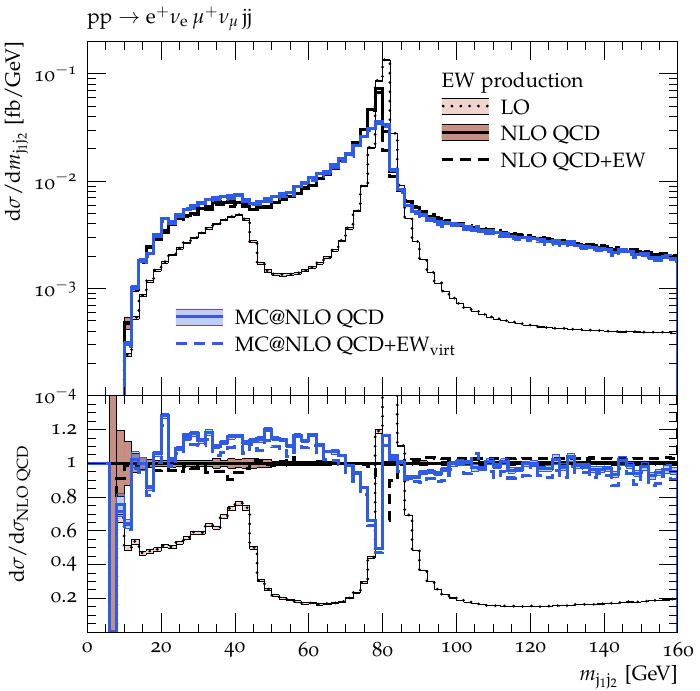}
\caption{Differential distributions for state-of-the-art predictions for the QCD (left) and EW (right) production for the ${\rm W}^+{\rm W}^+{\rm j}{\rm j}$ signature.
The distributions are taken from Ref.~\cite{Denner:2024ufg}.}
\label{fig:PS}
\end{figure}

Along with a fixed-order analysis, state-of-the-art predictions featuring parton-shower (PS) corrections using the public Monte Carlo program {\sc Sherpa}~\cite{Sherpa:2019gpd} has been presented for both the EW and the QCD production as illustrated in Fig.~\ref{fig:PS}.
The parton-shower corrections are about $10\%$ for the fiducial cross section but can be significantly more pronounced for some differential distributions.

\section{HH production}

The double-Higgs production is certainly one of the next processes to be measured at the LHC.
It is particularly crucial for the exploration of the Electroweak Symmetry Breaking Mechanism by providing sensitivity to triple-Higgs coupling.

Given that the partonic process ${\rm g}{\rm g} \to {\rm H}{\rm H}$ is loop-induced, 
LO predictions are already a one-loop calculation while the next-to-leading oder (NLO) predictions feature two-loop contributions.
For the EW corrections, the real radiation, which is a photon radiation, is actually vanishing due to Furry's theorem.
Note that weak radiations (W,Z) are not needed for infrared finiteness of the cross section and are typically rejected in experimental analysis.
It also means that the calculation of the NLO EW corrections only contain the two-loop virtual contribution.

This particularly challenging calculation has been tackled by several independent groups.
In Ref.~\cite{Bi:2023bnq} the exact calculation was presented while in Refs.~\cite{Borowka:2018pxx,Davies:2022ram,Muhlleitner:2022ijf,Davies:2023npk,Heinrich:2024dnz} several partial results and/or expanded results have been made public.

\begin{figure}
\includegraphics[width=0.49\linewidth]{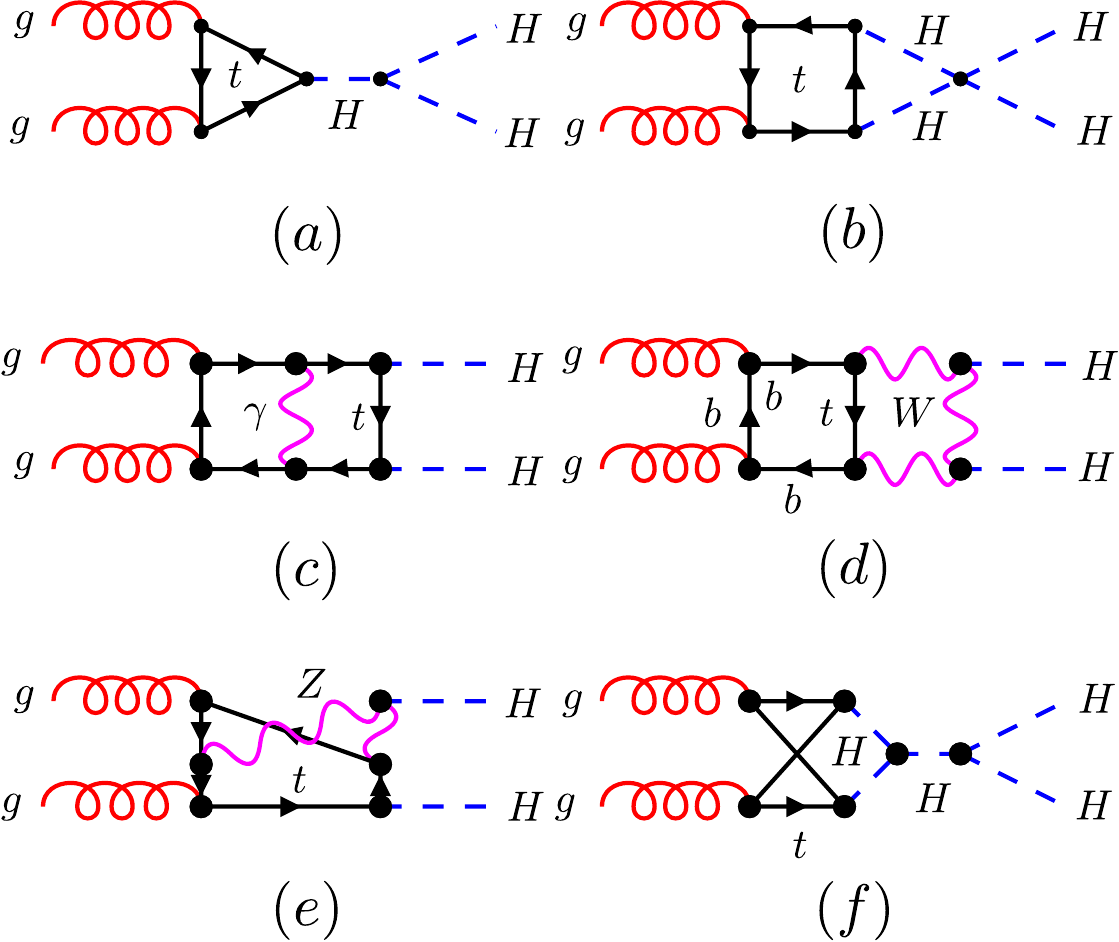}
\hfill
\raisebox{0.4cm}{\includegraphics[width=0.49\linewidth]{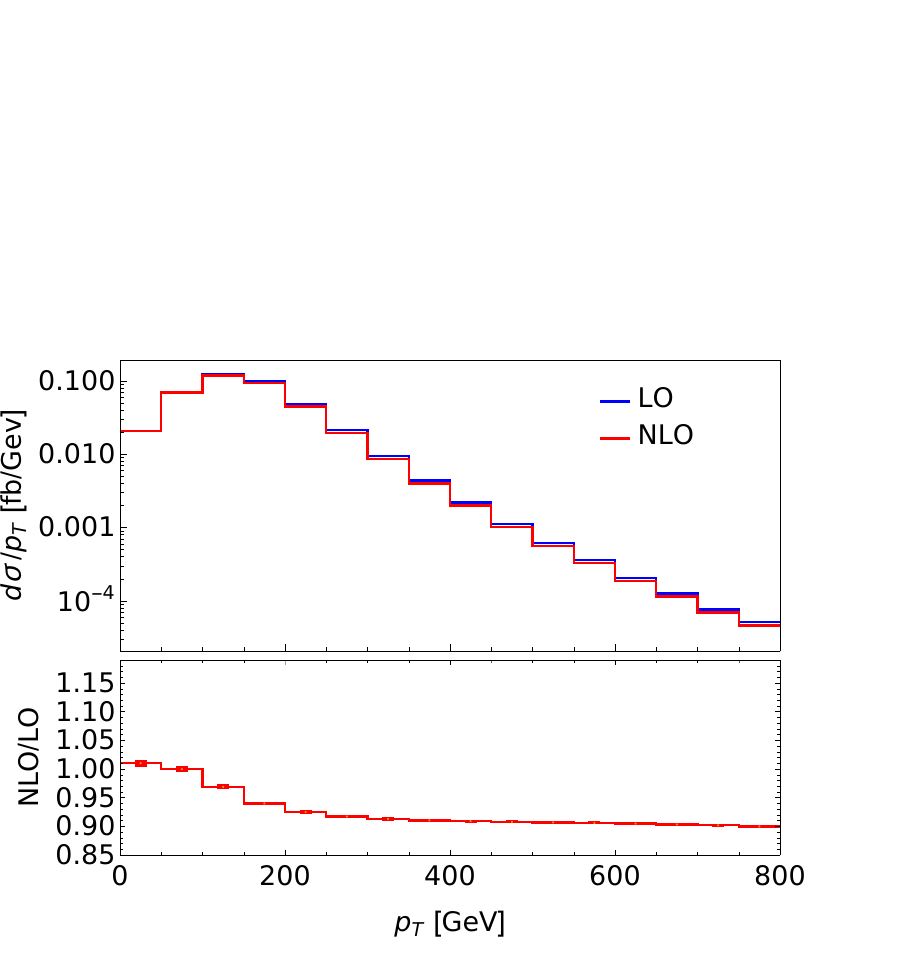}}
\caption{Sample Feynman diagram for ${\rm g}{\rm g} \to {\rm H}{\rm H}$ at the LHC at LO and NLO EW (left).
Distribution in the transverse momentum of one of the two Higgs bosons (right).
All figures are taken from Ref.~\cite{Bi:2023bnq}.}
\label{fig:gghh}
\end{figure}

Example contributing Feynman diagrams are shown on the left hand-side of Fig.~\ref{fig:gghh}.
The top diagrams display the LO contributions while the others show the two-loop EW virtual contributions.
On the right-hand side of Fig.~\ref{fig:gghh}, the LO and NLO EW predictions for the full calculation of Ref.~\cite{Bi:2023bnq} are shown.
The EW corrections display a typical behaviour for the LHC, namely that the corrections grow negatively large in the high-energy limit under the influence of Sudakov logarithms.
For example, the corrections are at the level of $-10\%$ at $800$ GeV.

\section{ttH production}

The production of a Higgs boson along with a top-antitop pair is of particular importance as it offers a key handle on the coupling between the Higgs boson and the top quark.

\begin{figure}
\begin{center}
\includegraphics[width=0.49\linewidth]{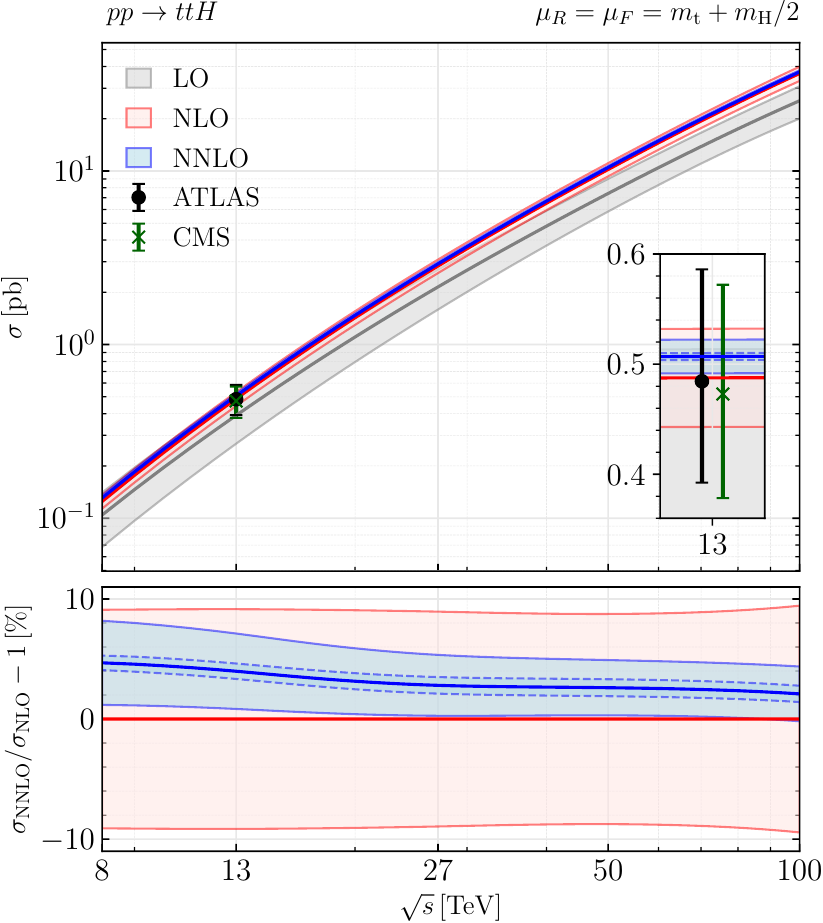} 
\end{center}
\caption{Inclusive cross sections as a function of the centre-of-mass energy in perturbative QCD for ttH at the LHC.
The figure is taken from Ref.~\cite{Catani:2022mfv}.}
\label{fig:tth}
\end{figure}

In Ref.~\cite{Catani:2022mfv}, a full computation at next-to-next-to-leading order (NNLO) QCD accuracy was presented.
The only approximation made in the calculation is in the two-loop contribution, which is the most challenging part of this calculation.
For the two-loop virtual contribution, the authors reverted to the so-called \emph{soft Higgs} approximation which is valid for scales much larger than the transverse momentum of the Higgs boson.
Such an approximation is appropriate for inclusive numbers and has been found to be $1\%$ accurate in this case.
In Fig.~\ref{fig:tth}, inclusive cross sections as a function of the centre-of-mass energy at the LHC are displayed at NNLO QCD accuracy.
The corrections are moderate, being at the level of $+4\%$ at $13$ TeV and $+2\%$ at $100$ TeV for example.

\begin{figure}
\begin{minipage}{0.49\textwidth}
 \includegraphics[width=0.49\linewidth]{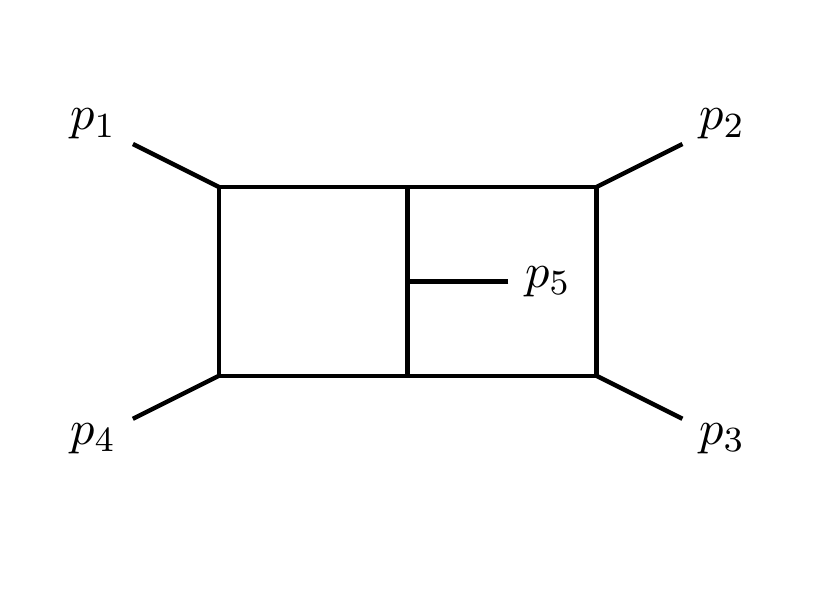}
 \includegraphics[width=0.49\linewidth]{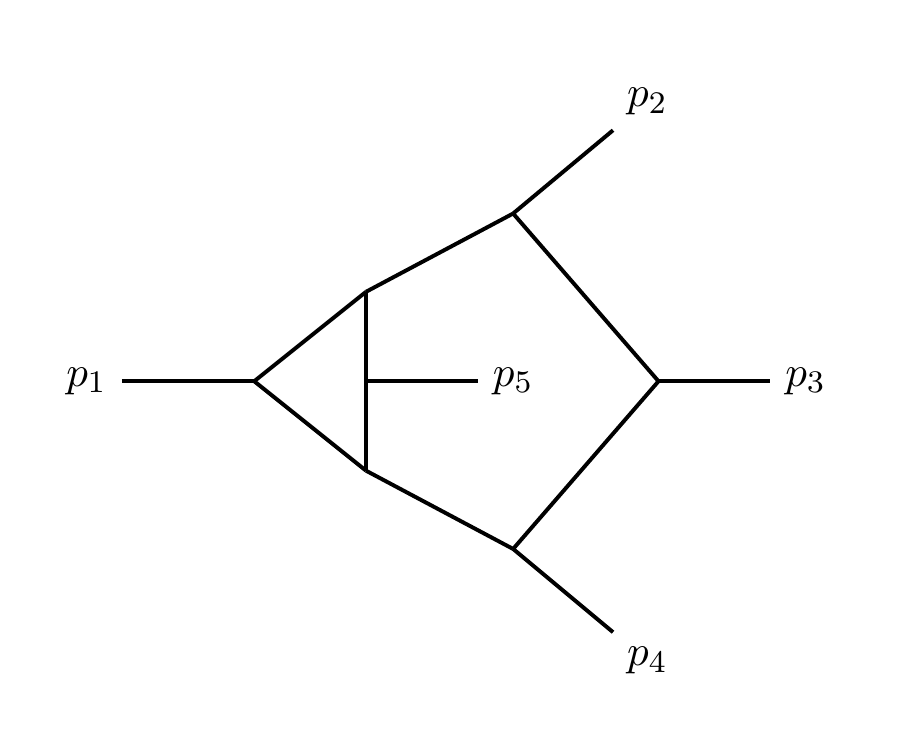}
 \includegraphics[width=0.49\linewidth]{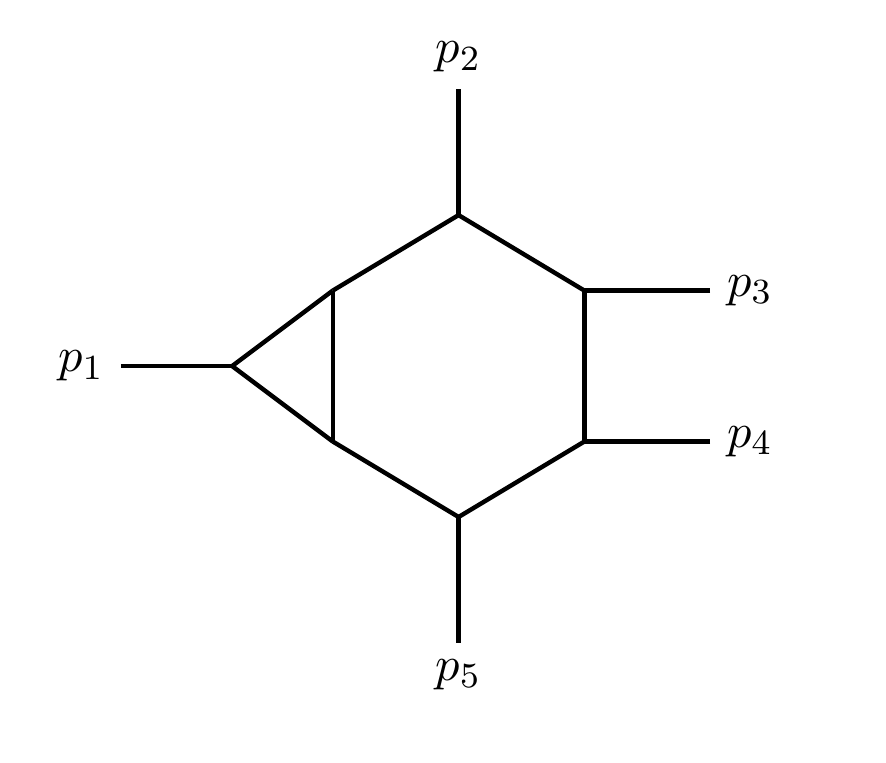}
 \includegraphics[width=0.49\linewidth]{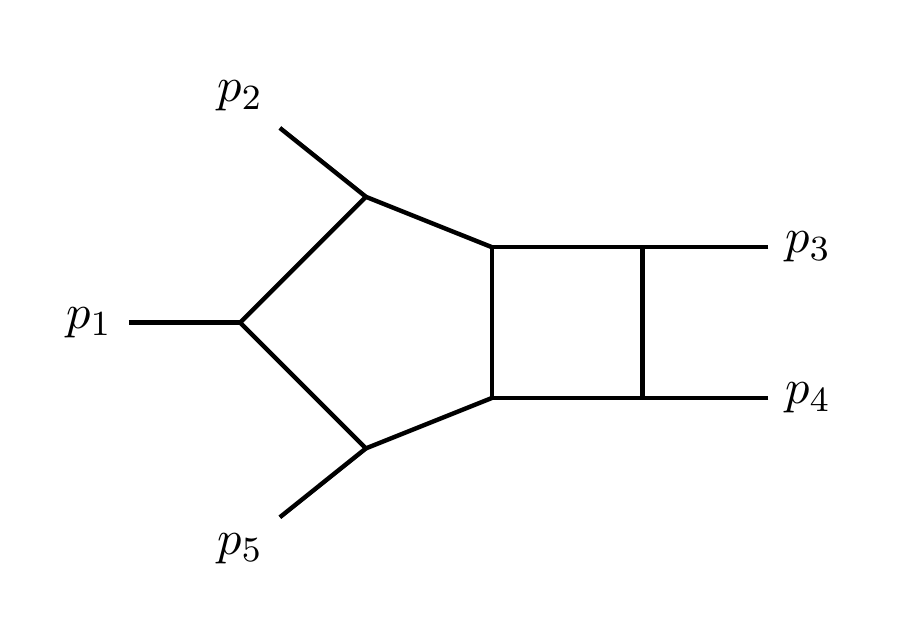}
\end{minipage}
\hfill
\begin{minipage}{0.49\textwidth}
\includegraphics[width=\linewidth]{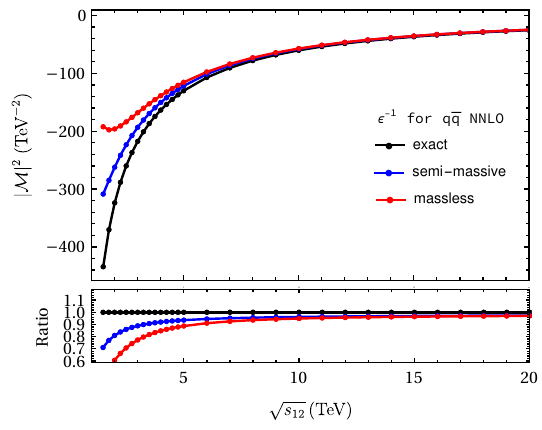}
\end{minipage}
\caption{Topologies contributing to the two-loop amplitude for ttH production at the LHC (left). Example numerical results for squared amplitudes at $\mathcal{O}\!\left(\epsilon^{-1}\right)$ for NNLO QCD for the $q\bar q$ channel (right).
All figures are taken from Ref.~\cite{Wang:2024pmv}.}
\label{fig:tth_2L}
\end{figure}

As mentioned before, the two-loop contributions are particularly challenging as they require the computation of 5-point two-loop amplitudes with 3 external masses and two different scales.
The topologies appearing in the computation are displayed in Fig.~\ref{fig:tth_2L} on the left-hand side.
In Ref.~\cite{Wang:2024pmv}, another approximation for both the $q\bar q$ and ${\rm g}{\rm g}$ channels has been presented.
This one is valid for $s_{ij}\gg m^2_{\rm t}$ i.e.\ in boosted topologies.
Exemplary numerical results are shown on the right hand-side of Fig.~\ref{fig:tth_2L}.
As pointed out by the authors, it would be interesting to combine the two approximations presented above.
The combined result should provide a rather reliable proxy for the two-loop corrections valid over the full phase space.

\begin{figure}
\includegraphics[width=0.49\linewidth]{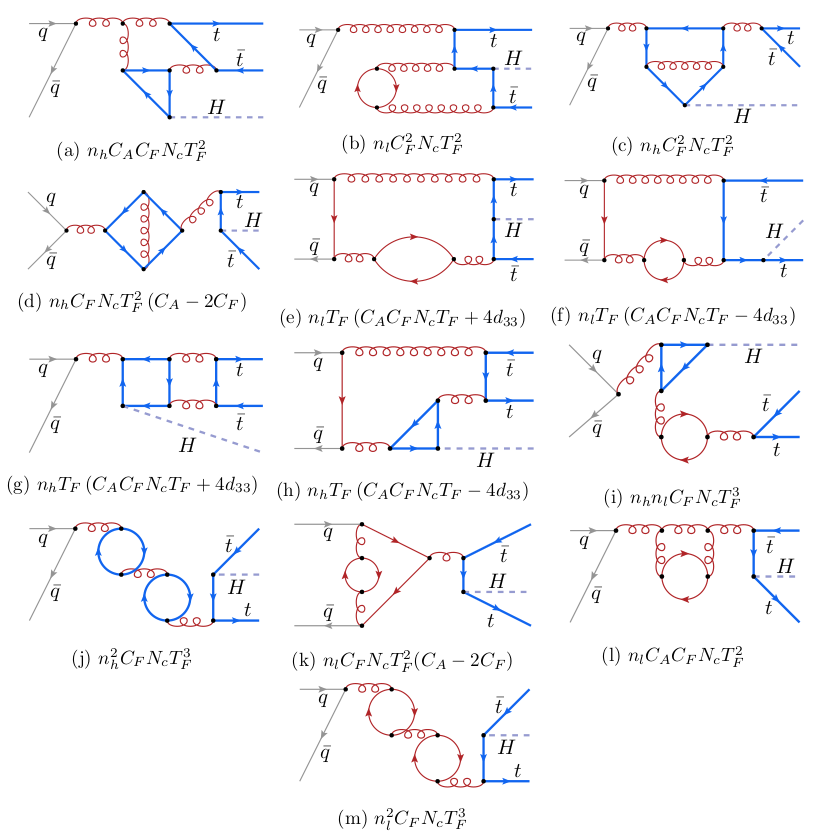}
\hfill
\includegraphics[width=0.49\linewidth]{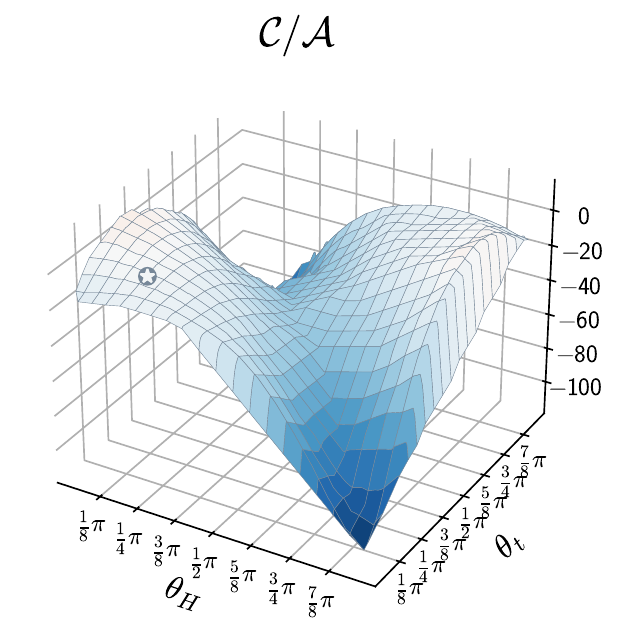}
\caption{
Sample Feynman diagrams contributing to the $N_f$ part of the two-loop QCD amplitude of ttH in the $q\bar q$ channel (left).
Numerical results for the two-loop virtual amplitudes (right).
All figures are taken from Ref.~\cite{Agarwal:2024jyq}.}
\label{fig:tth_num}
\end{figure}

Finally, it is worth mentioning Ref.~\cite{Agarwal:2024jyq} where the $N_f$ part of the two-loop amplitude (i.e.\ with closed fermion loops) of the $q\bar q$ channel has been computed.
Some contributing Feynman diagrams are provided on the left hand-side of Fig.~\ref{fig:tth_num}.
It is important to point out that, as opposed to the two previous cases, this calculation provides exact numerical results (as illustrated for example on the right hand-side of Fig.~\ref{fig:tth_num}).
It constitutes a proof of concept for the full calculation.
Note that the drawback of this approach is that the computation of each phase-space point takes several minutes.
This means that in order to use such results for Monte Carlo integration, interpolation grids will be needed.
Finally, note that progress towards a calculation of two-loop virtual contributions without approximation by other groups has been started~\cite{FebresCordero:2023pww,Buccioni:2023okz}.

\section{ttW production}

The production of a top-antitop pair in association with a W boson is interesting in its own right as it is a typical background for beyond-the-Standard Model signature.
In addition, it is also an important background to ttH production.

\begin{figure}
  \includegraphics[width=0.24\linewidth]{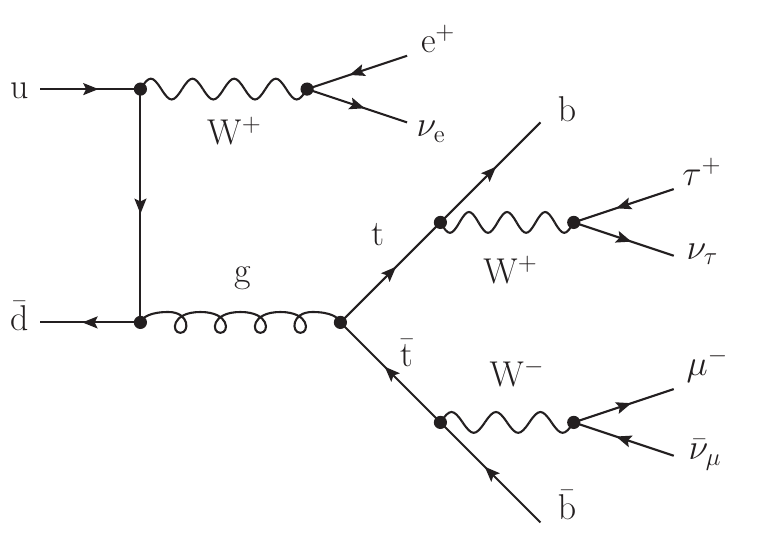}
  \includegraphics[width=0.24\linewidth]{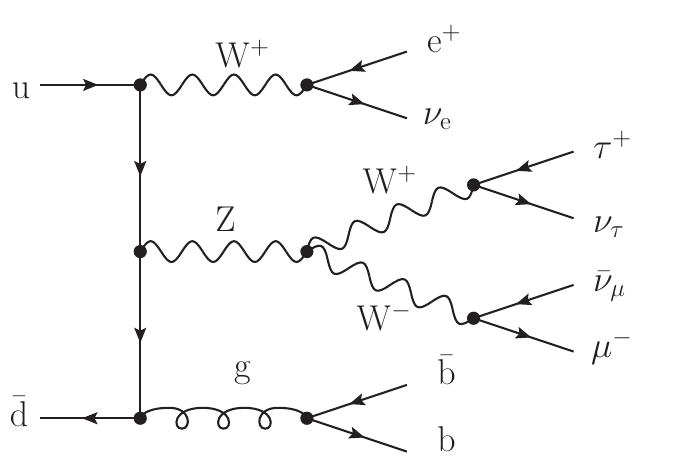}
  \includegraphics[width=0.24\linewidth]{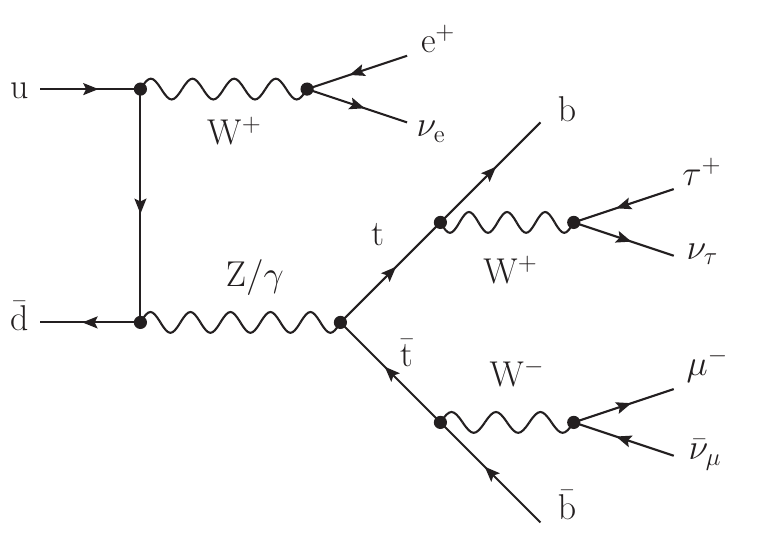}
  \includegraphics[width=0.24\linewidth]{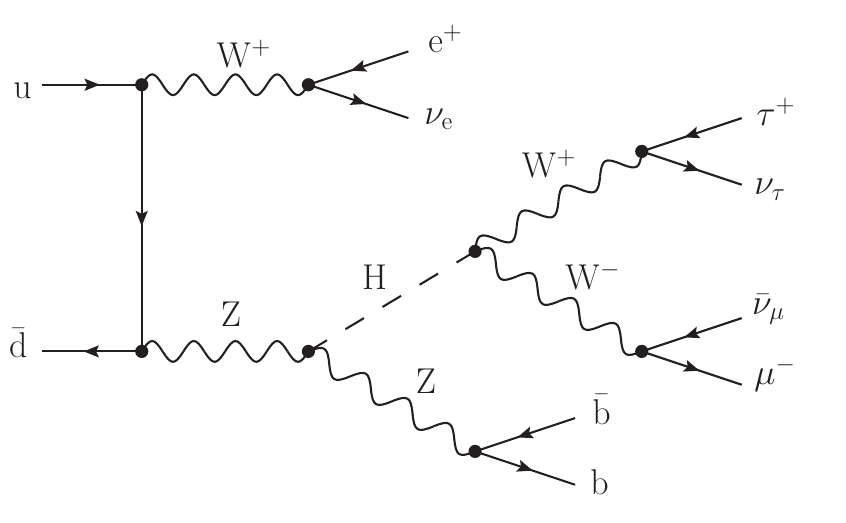}
\caption{
Sample Feynman diagrams contributing to the ttW final state for both the EW and QCD production featuring doubly-top resonant and non-top resonant contributions.
All diagrams are taken from Ref.~\cite{Denner:2021hqi}.}
\label{fig:ttW}
\end{figure}

In Ref.~\cite{Denner:2021hqi}, the full NLO QCD + EW corrections have been computed for the off-shell process.
This involves calculating corrections (QCD and EW) to both the EW and the QCD production process.
Representative tree-level diagrams are displayed in Fig.~\ref{fig:ttW}.
This calculation is therefore for a $2\to8$ process and hence particularly challenging.
One interesting aspect of the NLO corrections is that it has been observed that subleading corrections which are denoted by NLO3 (QCD corrections to the EW production) are particularly large~\cite{Dror:2015nkp,Frederix:2017wme} for on-shell top quarks and a W boson.
Following power-counting arguments, one expects this correction to be rather suppressed.
It is actually not the case due to ${\rm W}{\rm t}\to{\rm W}{\rm t}$ scattering arising at this order.

\begin{figure}[h]
  \includegraphics[width=0.49\linewidth]{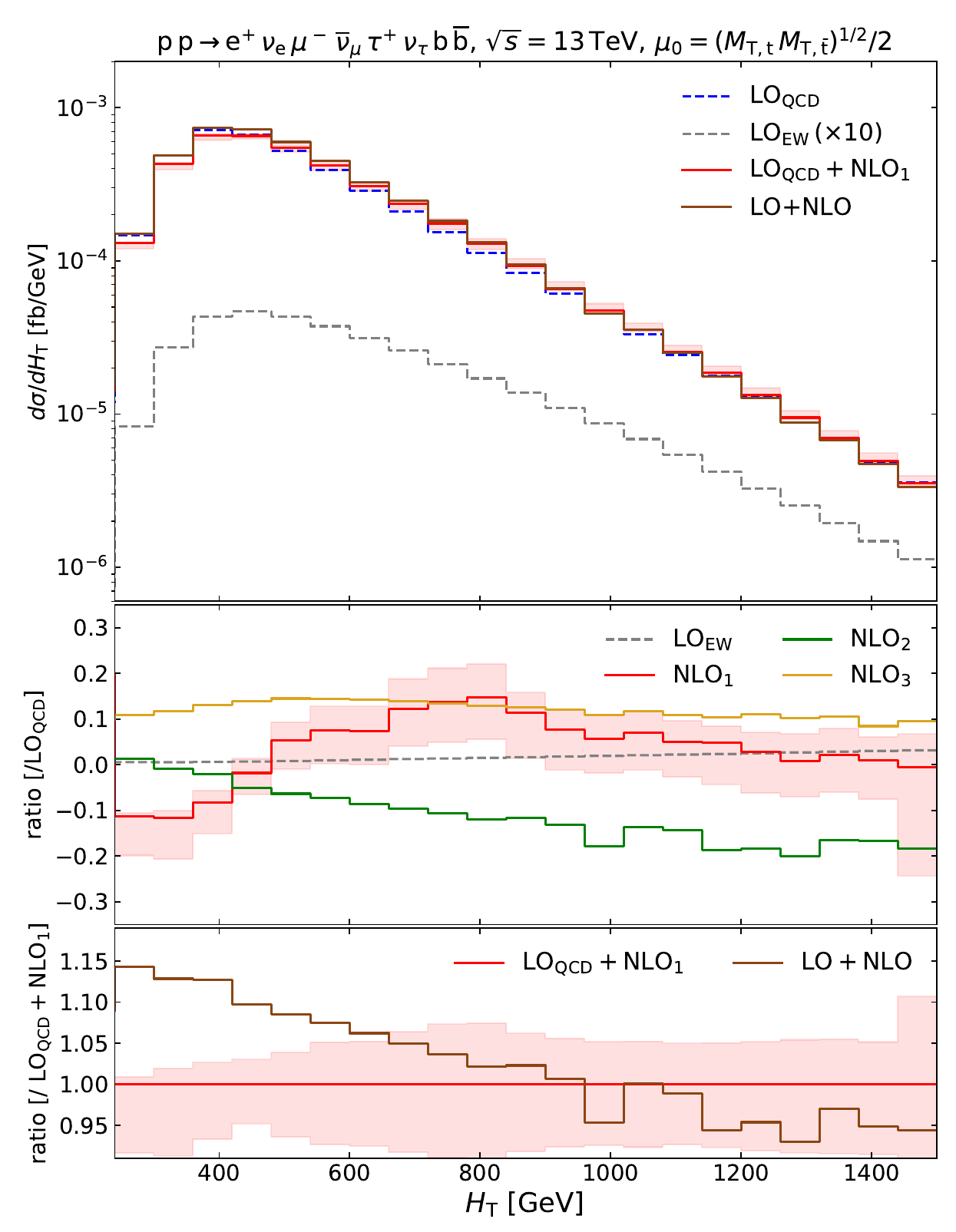}
  \hfill
  \includegraphics[width=0.49\linewidth]{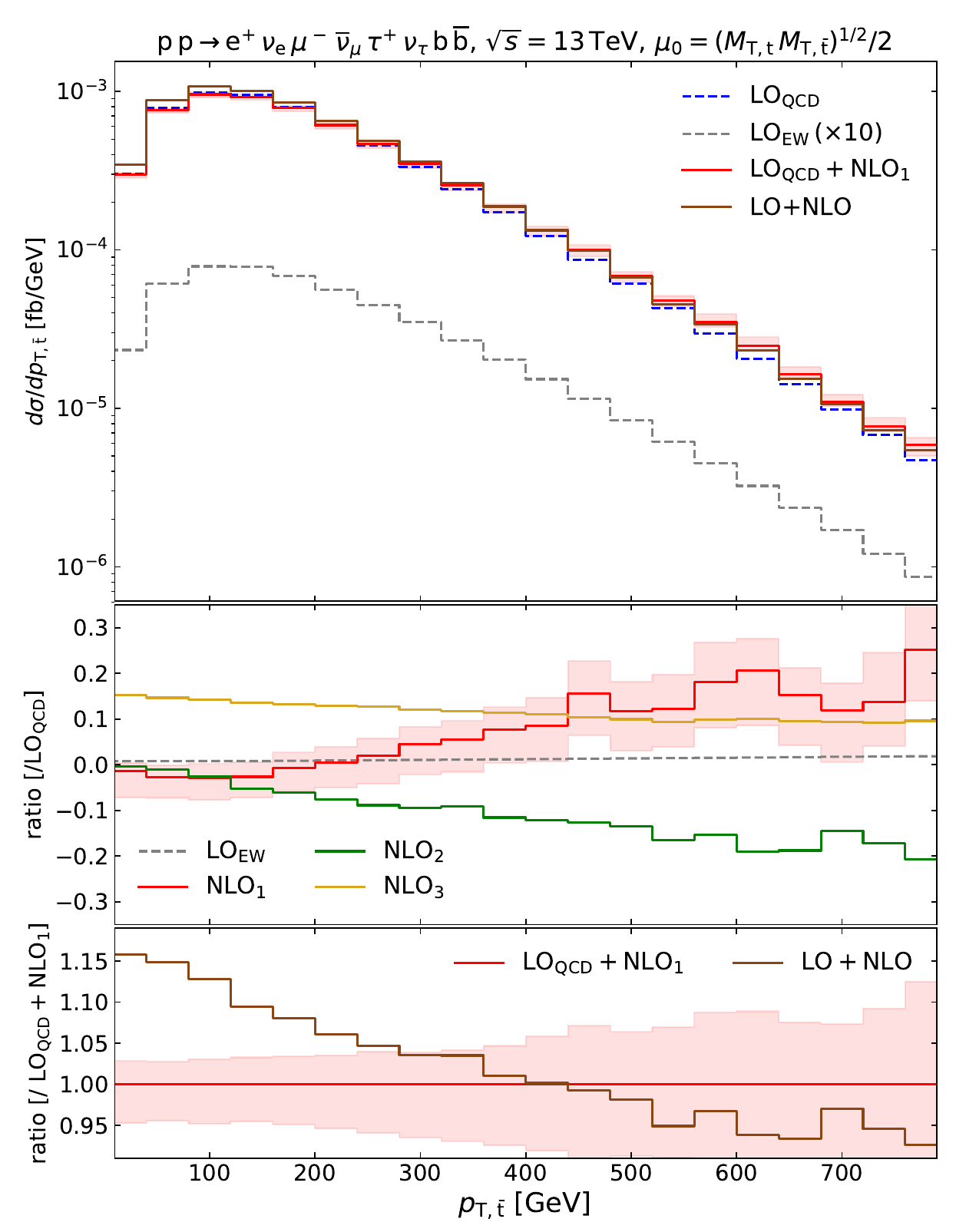}
\caption{
Differential distributions for off-shell ttW at the LHC a full NLO accuracy.
All diagrams are taken from Ref.~\cite{Denner:2021hqi}.}
\label{fig:ttW_plot}
\end{figure}

Reference~\cite{Denner:2021hqi} confirmed this finding for the full off-shell calculation.
In Fig.~\ref{fig:ttW_plot}, two differential distributions are displayed.
It shows that apart from the NLO4 contribution (EW corrections to the EW production), all other three corrections are actually phenomenologically relevant.
In addition, one can observe that across the whole phase space, the corrections display a non-trivial interplay, again preventing from neglecting any of them.

\section{Conclusion}

In these proceedings, I have highlighted recent computations that illustrate the current frontier of theoretical predictions at the LHC within the Standard Model.
The interested reader is referred to the original work for more information.
The processes mentioned above were all related to some extent to Higgs-boson physics due to its prevalence at the LHC.
In detail, the list of the processes reported with their corresponding features reads
 \begin{itemize}
 \item ${\rm p}{\rm p}\to{\rm W}{\rm W}{\rm W}/{\rm W}{\rm H}$ \hspace{1.7cm} [Full NLO QCD+EW with PS for $2\to6$]
 \item ${\rm p}{\rm p}\to{\rm H}{\rm H}$ \hspace{3.05cm} [Two-loop EW virtual]
 \item ${\rm p}{\rm p}\to{\rm t}\bar{\rm t}{\rm H}$ \hspace{3.2cm}[Approximate NNLO QCD/two-loop QCD virtual for $2\to3$]
 \item ${\rm p}{\rm p}\to{\rm t}\bar{\rm t}{\rm W}$  \hspace{3.0cm} [Full NLO QCD+EW for $2\to8$]
 \end{itemize}

These calculations constitute corner stones of the precision programme currently ongoing at the LHC.
They are therefore decisive information for Standard Model tests that will be performed during the run~III of the LHC and beyond.

\acknowledgments

MP acknowledges support by the German Research Foundation (DFG) through the Research Training Group RTG2044.

% \newpage


\begin{thebibliography}{99}
%\cite{Jakobs:2023fxh}
\bibitem{Jakobs:2023fxh}
K.~Jakobs and G.~Zanderighi,
``The profile of the Higgs boson: status and prospects,''
Phil. Trans. Roy. Soc. Lond. A \textbf{382} (2023) no.2266, 20230087
doi:10.1098/rsta.2023.0087
[arXiv:2311.10346 [hep-ph]].
%2 citations counted in INSPIRE as of 01 Jul 2024
%\cite{Huss:2022ful}
\bibitem{Huss:2022ful}
A.~Huss, J.~Huston, S.~Jones and M.~Pellen,
``Les Houches 2021\textemdash{}physics at TeV colliders: report on the standard model precision wishlist,''
J. Phys. G \textbf{50} (2023) no.4, 043001
doi:10.1088/1361-6471/acbaec
[arXiv:2207.02122 [hep-ph]].
%28 citations counted in INSPIRE as of 01 Jul 2024
%\cite{Andersen:2024czj}
\bibitem{Andersen:2024czj}
J.~Andersen, B.~Assi, K.~Asteriadis, P.~Azzurri, G.~Barone, A.~Behring, A.~Benecke, S.~Bhattacharya, E.~Bothmann and S.~Caletti, \textit{et al.}
``Les Houches 2023: Physics at TeV Colliders: Standard Model Working Group Report,
[arXiv:2406.00708 [hep-ph]].
%3 citations counted in INSPIRE as of 01 Jul 2024
%\cite{Jones:2023uzh}
\bibitem{Jones:2023uzh}
S.~P.~Jones,
``An Overview of Standard Model Calculations for Higgs Boson Production \& Decay,''
LHEP \textbf{2023} (2023), 442
doi:10.31526/lhep.2023.442
%2 citations counted in INSPIRE as of 01 Jul 2024
%\cite{Biedermann:2017bss}
\bibitem{Biedermann:2017bss}
B.~Biedermann, A.~Denner and M.~Pellen,
``Complete NLO corrections to W$^{+}$W$^{+}$ scattering and its irreducible background at the LHC,''
JHEP \textbf{10} (2017), 124
doi:10.1007/JHEP10(2017)124
[arXiv:1708.00268 [hep-ph]].
%102 citations counted in INSPIRE as of 01 Jul 2024
\bibitem{Denner:2024ufg}
A.~Denner, M.~Pellen, M.~Sch\"onherr and S.~Schumann,
``Tri-boson and WH production in the $\mathrm{W}^+\mathrm{W}^+\mathrm{j}\mathrm{j}$ channel: predictions at full NLO accuracy and beyond,''
[arXiv:2406.11516 [hep-ph]].
%0 citations counted in INSPIRE as of 01 Jul 2024
%\cite{ATLAS:2022xnu}
\bibitem{ATLAS:2022xnu}
G.~Aad \textit{et al.} [ATLAS],
``Observation of $WWW$ Production in $pp$ Collisions at $\sqrt s$ =13\,\,TeV with the ATLAS Detector,''
Phys. Rev. Lett. \textbf{129} (2022) no.6, 061803
doi:10.1103/PhysRevLett.129.061803
[arXiv:2201.13045 [hep-ex]].
%34 citations counted in INSPIRE as of 01 Jul 2024
%\cite{Denner:2024ufg}
%\cite{Biedermann:2016yds}
\bibitem{Biedermann:2016yds}
B.~Biedermann, A.~Denner and M.~Pellen,
``Large electroweak corrections to vector-boson scattering at the Large Hadron Collider,''
Phys. Rev. Lett. \textbf{118} (2017) no.26, 261801
doi:10.1103/PhysRevLett.118.261801
[arXiv:1611.02951 [hep-ph]].
%90 citations counted in INSPIRE as of 01 Jul 2024
%\cite{Sherpa:2019gpd}
\bibitem{Sherpa:2019gpd}
E.~Bothmann \textit{et al.} [Sherpa],
``Event Generation with Sherpa 2.2,''
SciPost Phys. \textbf{7} (2019) no.3, 034
doi:10.21468/SciPostPhys.7.3.034
[arXiv:1905.09127 [hep-ph]].
%1013 citations counted in INSPIRE as of 01 Jul 2024
%\cite{Bi:2023bnq}
\bibitem{Bi:2023bnq}
H.~Y.~Bi, L.~H.~Huang, R.~J.~Huang, Y.~Q.~Ma and H.~M.~Yu,
``Electroweak Corrections to Double Higgs Production at the LHC,''
Phys. Rev. Lett. \textbf{132} (2024) no.23, 231802
doi:10.1103/PhysRevLett.132.231802
[arXiv:2311.16963 [hep-ph]].
%9 citations counted in INSPIRE as of 01 Jul 2024
%\cite{Borowka:2018pxx}
\bibitem{Borowka:2018pxx}
S.~Borowka, C.~Duhr, F.~Maltoni, D.~Pagani, A.~Shivaji and X.~Zhao,
%``Probing the scalar potential via double Higgs boson production at hadron colliders,''
JHEP \textbf{04}, 016 (2019)
doi:10.1007/JHEP04(2019)016
[arXiv:1811.12366 [hep-ph]].
%64 citations counted in INSPIRE as of 30 Jul 2024
%\cite{Davies:2022ram}
\bibitem{Davies:2022ram}
J.~Davies, G.~Mishima, K.~Sch\"onwald, M.~Steinhauser and H.~Zhang,
``Higgs boson contribution to the leading two-loop Yukawa corrections to gg \textrightarrow{} HH,''
JHEP \textbf{08} (2022), 259
doi:10.1007/JHEP08(2022)259
[arXiv:2207.02587 [hep-ph]].
%25 citations counted in INSPIRE as of 30 Jul 2024
%\cite{Muhlleitner:2022ijf}
\bibitem{Muhlleitner:2022ijf}
M.~M\"uhlleitner, J.~Schlenk and M.~Spira,
``Top-Yukawa-induced corrections to Higgs pair production,''
JHEP \textbf{10} (2022), 185
doi:10.1007/JHEP10(2022)185
[arXiv:2207.02524 [hep-ph]].
%19 citations counted in INSPIRE as of 30 Jul 2024
%\cite{Davies:2023npk}
\bibitem{Davies:2023npk}
J.~Davies, K.~Sch\"onwald, M.~Steinhauser and H.~Zhang,
``Next-to-leading order electroweak corrections to $gg \to HH$ and $gg \to gH$ in the large-$m_t$ limit,''
JHEP \textbf{10} (2023), 033
doi:10.1007/JHEP10(2023)033
[arXiv:2308.01355 [hep-ph]].
%14 citations counted in INSPIRE as of 01 Jul 2024
%\cite{Heinrich:2024dnz}
\bibitem{Heinrich:2024dnz}
G.~Heinrich, S.~Jones, M.~Kerner, T.~Stone and A.~Vestner,
``Electroweak corrections to Higgs boson pair production: The top-Yukawa and self-coupling contributions,''
[arXiv:2407.04653 [hep-ph]].
%0 citations counted in INSPIRE as of 08 Jul 2024
%\cite{Catani:2022mfv}
\bibitem{Catani:2022mfv}
S.~Catani, S.~Devoto, M.~Grazzini, S.~Kallweit, J.~Mazzitelli and C.~Savoini,
``Higgs Boson Production in Association with a Top-Antitop Quark Pair in Next-to-Next-to-Leading Order QCD,''
Phys. Rev. Lett. \textbf{130} (2023) no.11, 111902
doi:10.1103/PhysRevLett.130.111902
[arXiv:2210.07846 [hep-ph]].
%31 citations counted in INSPIRE as of 01 Jul 2024
%\cite{Wang:2024pmv}
\bibitem{Wang:2024pmv}
G.~Wang, T.~Xia, L.~L.~Yang and X.~Ye,
``Two-loop QCD amplitudes for $t\bar{t}H$ production from boosted limit,''
[arXiv:2402.00431 [hep-ph]].
%3 citations counted in INSPIRE as of 01 Jul 2024
%\cite{Agarwal:2024jyq}
\bibitem{Agarwal:2024jyq}
B.~Agarwal, G.~Heinrich, S.~P.~Jones, M.~Kerner, S.~Y.~Klein, J.~Lang, V.~Magerya and A.~Olsson,
``Two-loop amplitudes for $ t\overline{t}H $ production: the quark-initiated N$_{f}$-part,''
JHEP \textbf{05} (2024), 013
doi:10.1007/JHEP05(2024)013
[arXiv:2402.03301 [hep-ph]].
%3 citations counted in INSPIRE as of 01 Jul 2024
%\cite{FebresCordero:2023pww}
\bibitem{FebresCordero:2023pww}
F.~Febres Cordero, G.~Figueiredo, M.~Kraus, B.~Page and L.~Reina,
``Two-loop master integrals for leading-color $ pp\to t\overline{t}H $ amplitudes with a light-quark loop,''
JHEP \textbf{07} (2024), 084
doi:10.1007/JHEP07(2024)084
[arXiv:2312.08131 [hep-ph]].
%6 citations counted in INSPIRE as of 16 Jul 2024
%\cite{Buccioni:2023okz}
\bibitem{Buccioni:2023okz}
F.~Buccioni, P.~A.~Kreer, X.~Liu and L.~Tancredi,
``One loop QCD corrections to gg \textrightarrow{} $t\overline{t }H$ at $\mathcal{O}\left({\epsilon }^{2}\right)$,''
JHEP \textbf{03} (2024), 093
doi:10.1007/JHEP03(2024)093
[arXiv:2312.10015 [hep-ph]].
%2 citations counted in INSPIRE as of 16 Jul 2024
%\cite{Denner:2021hqi}
\bibitem{Denner:2021hqi}
A.~Denner and G.~Pelliccioli,
``Combined NLO EW and QCD corrections to off-shell $\text {t} \overline{\text {t}}\text {W} $ production at the LHC,''
Eur. Phys. J. C \textbf{81} (2021) no.4, 354
doi:10.1140/epjc/s10052-021-09143-3
[arXiv:2102.03246 [hep-ph]].
%33 citations counted in INSPIRE as of 01 Jul 2024
%\cite{Frederix:2017wme}
\bibitem{Frederix:2017wme}
R.~Frederix, D.~Pagani and M.~Zaro,
``Large NLO corrections in $t\bar{t}W^{\pm}$ and $t\bar{t}t\bar{t}$ hadroproduction from supposedly subleading EW contributions,''
JHEP \textbf{02} (2018), 031
doi:10.1007/JHEP02(2018)031
[arXiv:1711.02116 [hep-ph]].
%221 citations counted in INSPIRE as of 01 Jul 2024
%\cite{Dror:2015nkp}
\bibitem{Dror:2015nkp}
J.~A.~Dror, M.~Farina, E.~Salvioni and J.~Serra,
``Strong tW Scattering at the LHC,''
JHEP \textbf{01} (2016), 071
doi:10.1007/JHEP01(2016)071
[arXiv:1511.03674 [hep-ph]].
%99 citations counted in INSPIRE as of 01 Jul 2024

\end{thebibliography}
\end{document}